\documentstyle[12pt]{article}
\begin{document}


\begin{center} 
{\Large{\bf  Supersymmetry on a Lattice }}\\
\vskip 0.15cm
{\Large{\bf and }}\\
\vskip 0.15cm
{\Large{\bf Dirac Fermions in a Random Vector Potential }}
\vskip 1.5cm

{\Large  Ikuo Ichinose\footnote{e-mail 
 address: ikuo@hep1.c.u-tokyo.ac.jp}}  
\vskip 0.5cm
 
Institute of Physics, University of Tokyo, Komaba,
  Tokyo, 153-8902 Japan  
 
\end{center}

\vskip 2cm
\begin{center} 
\begin{bf}
Abstract
\end{bf}
\end{center}
We study two-dimensional Dirac fermions in a random non-Abelian 
vector potential by using lattice regularization.
We consider $U(N)$ random vector potential for large $N$. 
The ensemble average with respect to random vector potential 
is taken by using lattice supersymmetry which we introduced
before in order to investigate phase structure of supersymmetric
gauge theory.
We show that a phase transition occurs at a certain critical 
disorder strength.
The ground state and low-energy excitations are studied in detail
in the strong-disorder phase.
Correlation function of the fermion local density of states
decays algebraically at the band center because of a quasi-long-range 
order of chiral symmetry and the chiral anomaly cancellation
in the lattice regularization (the species doubling).
In the present study, we use the lattice regularization and also 
the Haar measure of $U(N)$ for the average over the random vector
potential.
Therefore topologically nontrivial configurations of the vector potential
are all included in the average.
Implication of the present results for the system of Dirac fermions in 
a random vector potential with noncompact Gaussian distribution
is discussed.

\newpage

\setcounter{footnote}{0}
\section{Introduction}

Random disordered systems are one of the most important
problems in condensed matter physics.
Especially in one and two dimensions effects of random disorders 
are so strong that almost all states are localized  
by random potentials.
Nonperturbative methods are required in order to investigate
random disordered systems in low dimensions.
Recently there appeared important studies on Dirac fermions
with random-varying mass and/or in a random vector potential.

Study of the Dirac fermions in a random vector potential was revived
in Ref.\cite{ludwig} in the context of quantum Hall plateau transition.
Non-Abelian generalization was introduced in Ref.\cite{nersesyan}
in the context of a d-wave superconductor.
After that, there appeared a number of interesting 
papers\cite{kogan,chamon1,chamon,mudry,castillo,caux}.

There are two technical problems for studying the system;  \\
(i) Normalization of states before the ensemble average over random
    variables.  \\
(ii) Integration over the white-noise random vector potential.  \\
For the first one, replica trick and supersymmetric(SUSY) methods
are often used.
The second problem is how to {\em regularize} the integral over 
the vector potential which has no spatial correlations.
Sometimes specific parameterization of vector potential
is used especially for the non-Abelian case\cite{mudry,caux}.
There only topologically trivial configurations are integrated over.
By using those ``technologies", critical lines in the system are
observed. 

Among various interesting properties of the random disordered systems, 
one of the most important problems is a disorder-induced
phase transition.
For the Dirac fermions in an Abelian random vector potential, 
a weak-strong disorder transition was found in Ref.\cite{chamon1} by 
studying multifractal scaling exponents of the critical wave function 
which is obtained exactly (see also Ref.\cite{castillo}).
Instability of the critical line is also seen by the existence
of an infinite set of operators with negative scaling 
dimensions\cite{ludwig,chamon,mudry} and very recently solution to the
negative dimension problem was suggested by Gurarie\cite{gurarie}.
For the case of non-Abelian random vector potential, existence
of operators with negative scaling dimensions was discussed in
Ref.\cite{mudry} and a termination mechanism was given in Ref.\cite{caux}
for the $SU(2)$ case.
As a related system with the Dirac fermion in a random vector potential,
random XY model in two dimensions was studied in Refs.\cite{XY1,XY2}.

In all the above previous studies, integral region of the vector potential
$A_\mu$ is {\em noncompact}, i.e., $A_\mu \in (-\infty, + \infty)$,  
and the probability distribution for it is taken to be Gaussian.
However in the network models by Chalker and Coddington\cite{network}
and their field-theory representation\cite{lee}, the random vector 
potential corresponds to the random Aharonov-Bohm phase for electron
which moves in a random potential.
Therefore the random variables are $e^{iA_\mu}\in U(1)$, and 
the range of the vector potential is {\em compact}, i.e.,
$A_\mu \in [-\pi, + \pi]$.
In the weak-disorder case, the above compact distribution of the 
vector potential might be approximated by the noncompact 
Gaussian distribution,
but at the strong disorder substantial differences in the properties
of the system are expected to appear. 

In this paper, we shall study random Dirac 
fermions by employing a lattice
regularization to define the system without any ambiguity.
The two-dimensional(2D) Dirac fermions in a random vector field 
is formulated on the 
lattice and the integral measure of  the vector potential is {\em compact}
as in the original network models\cite{network,lee}. 
Therefore topologically nontrivial configurations of 
the vector potential are all included.
We consider $U(N)$ vector potential for large $N$.
Because of the compactness of the group $U(N)$, 
the one-link integral is evaluated exactly for large $N$ in a closed form.
Concerning with the above problem (i), we shall use the lattice
SUSY(LSUSY) methods which we introduced before for the investigation
on the SUSY gauge theory\cite{ichinose1}.
Formulation of the LSUSY is an important but still unsolved problem.
However the LSUSY given in Ref.\cite{ichinose1} is suitable for the 
present study.

For the one-link integral over $U(N)$ for lage $N$,
it is known that there are two regimes
or ``phases", the weak-coupling and strong-coupling regimes.
More precisely in the present context, the one-link integral 
exhibits a third-order phase transition as 
the disorder strength is increased.
Then one can expect that there is a genuine phase transition corresponding
to these two regimes in the thermodynamic limit.
In this paper we study the system in both the weak and strong-disorder
cases and investigate the properties of both ``phases".
Especially as the strong-disorder limit acquires lots of interest recently,
we study the strong-disorder phase in detail.

This paper is organized as follows.
In Sect.2, the model and the LSUSY are explained.
In Sect.3, the weak-disorder regime is studied and an effective
action is obtained by integrating over the random vector potential.
In Sect.4, the strong-disorder regime is considered.
Properties of the ground state and low-energy excitations are
clarified.
It is shown that there exists a quasi-long-range order and 
correlation of the local density of states decays algebraically.
Section 5 is devoted for discussion.
Implication of the result for other interesting cases is discussed.
Especially relation between the properties of the present model and 
results of the previous studies is examined.
Physical picture of the result is explained.

\setcounter{equation}{0}
\section{Model and LSUSY}

We shall study 2D Dirac fermions in a $U(N)$ random vector potential
by employing the lattice regularization.
Action of the Dirac fermion $\psi^a$ ($a=1,..., N$) on the square lattice
is given by
\begin{equation}
S_D={1\over 2}\sum \Big[\bar{\psi}(x)\gamma_\mu U_\mu(x)\psi(x+\mu)
-\bar{\psi}(x+\mu)\gamma_\mu U^{\dagger}_\mu(x)\psi(x)\Big],
\label{SD}
\end{equation}
where $x=(x_0,x_1)$ denotes lattice site, $\mu=(0,1)$ is the direction index, 
$U_\mu(x)$ is $U(N)$ field on the link $(x,x+\mu)$ 
$U_\mu(x)=\Big(U_\mu(x)\Big)^a_b \in U(N)$ and we set the lattice spacing 
$a_L=1$.
The random $U(N)$ vector potential $A_{\mu,\alpha}(x)$ ($\alpha$ is the 
$U(N)$ index) is related with $U_\mu(x)$ as follows,
\begin{equation}
U_\mu(x)=e^{ia_L\sum_\alpha T^\alpha A_{\mu,\alpha}(x)},
\end{equation}
where $T^\alpha$'s are generators of the $U(N)=U(1)\times SU(N)$ Lie algebra.
In the (naive) continuum limit $a_L \rightarrow 0$, we can expand 
$U_\mu(x)$ as $U_\mu(x)=1+ia_L\sum_\alpha T^\alpha A_{\mu,\alpha}(x)
+\cdots$ and recover the usual action of the Dirac fermion in the continuum.
The two-dimensional $\gamma$-matrices are explicitly given by the 
Pauli matrices as $\gamma_0=\sigma_x,\; \gamma_1=\sigma_y$ and 
$\gamma_5=\sigma_z$.

In order to take the ensemble average over the vector potential 
as random variables, we shall
introduce boson field in a SUSY manner.
To this end we shall slightly rewrite $S_D$ in Eq.(\ref{SD}).
By the following transformation,
\begin{equation}
\psi(x)=T(x)\chi(x), \;\; \bar{\psi}(x)=\bar{\chi}(x)T^\dagger(x),
\label{transf}
\end{equation}
with $T(x)=(\gamma_0)^{x_0}(\gamma_1)^{x_1}$ and using the identities
like $(\gamma_\mu)^2=1$ and $(\gamma_0)^n\gamma_1=(-)^n\gamma_1(\gamma_0)^n$
($n$ is an integer), 
\begin{eqnarray}
S_D&=& {1\over 2}\sum \Big[\bar{\chi}(x)\eta_\mu(x)U_\mu(x)
\chi(x+\mu)-\bar{\chi}(x+\mu)\eta_\mu(x)
U^\dagger_\mu(x)\chi(x)\Big]  \nonumber   \\
&=&\sum\bar{\chi}(x)\hat{D}\chi(x),
\label{Schi}
\end{eqnarray}
where $\eta_0(x)=1, \; \eta_1(x)=(-)^{x_0}$, and 
\begin{equation}
\hat{D}\chi(x)={1\over 2}\sum_\mu\Big[\eta_\mu(x)U_\mu(x)
\chi(x+\mu)-\eta_\mu(x-\mu)U^\dagger_\mu(x-\mu)\chi(x-\mu)\Big].
\label{hatD}
\end{equation}
The fields $\chi$ and $\bar{\chi}$ are two-component spinors but
their spinor indices are diagonal in the action (\ref{Schi}).
We add the following ``mass term" to the action which measures deviation from
the band center or critical line,
\begin{eqnarray}
S_M&=&M\sum \bar{\chi}(x)\gamma_5\chi(x) \nonumber  \\
&=&M\sum\Big[\bar{\chi}_-\chi_+-\bar{\chi}_+\chi_-\Big],
\label{SM}
\end{eqnarray}
where $\chi=(\chi_+,\chi_-)^t$ and $\bar{\chi}=(\bar{\chi}_-,
\bar{\chi}_+)^t$.
The specific form of the above mass term comes from merely
a technical reason which becomes clear shortly.
We are interested in the limit $M \rightarrow 0$.

We introduce a complex scalar field $\phi(x)$ whose action is
given by
\begin{equation}
S_\phi=\sum\hat{D}\phi^\dagger(x)\hat{D}\phi(x)+
m^2\sum \phi^\dagger(x)\phi(x).
\label{Sphi}
\end{equation}
As the modified ``Dirac" operator $\hat{D}$ does not contain
the $\gamma$-matrices, the same $\hat{D}$ can be applied for the 
scalar field $\phi$.
This is an essential point in the present construction of
the SUSY lattice model.   
It can be shown that in the classical continuum limit
the action $S_\phi$ in Eq.(\ref{Sphi}) reduces to the 
usual action of the scalar field in the continuum,
and the integral over the scalar fields is well-defined
because the action is positive-definite. 

The total action of the system is given by
\begin{eqnarray}
S&=&S_\chi+S_\phi,  \nonumber  \\
S_\chi&=&S_D+S_M  \nonumber   \\
&=&\sum \Big[\bar{\chi}_+\hat{D}\chi_-+\bar{\chi}_-\hat{D}
\chi_+\Big]+M\sum\Big[\bar{\chi}_-\chi_+-\bar{\chi}_+\chi_-\Big].
\label{ST}
\end{eqnarray}
It is seen that the partition function is just unity if $M=m$
for an arbitrary fixed configuration of the vector potential
because of the cancellation of the fermion and boson determinants.
Actually
\begin{eqnarray}
 \int [D\phi^\dagger D\phi]\; e^{-S_\phi}
&=&\mbox{det}^{-1}(-\hat{D}^2+m^2), \nonumber  \\
 \int [D\bar{\chi}D\chi]\; e^{-S_\chi}
&=&\mbox{Det}\left(
           \begin{array}{cc}
           \hat{D}+M & 0 \\
           0 & \hat{D}-M 
           \end{array}
           \right)    \nonumber  \\
&=&\mbox{det}(\hat{D}^2-M^2),
\end{eqnarray}
where Det is the determinant of the spinor and real spaces and 
det is that of the real space.

Moreover the action $S$ is invariant under the following
LSUSY transformation for $M=m$,
\begin{eqnarray}
\delta\phi&=&\bar{\epsilon}_+\chi_-+\bar{\epsilon}_-\chi_+, \nonumber  \\
\delta\phi^\dagger&=&\bar{\chi}_-\epsilon_++\bar{\chi}_+
\epsilon_-,   \nonumber  \\
\delta\chi_{\pm}&=&\pm M\phi\epsilon_{\pm}-
\hat{D}\phi\epsilon_{\pm},  \nonumber  \\
\delta \bar{\chi}_{\pm}&=&\pm M\phi^\dagger\bar{\epsilon}_\pm
-\hat{D}\phi^\dagger\bar{\epsilon}_\pm,
\label{SUSY}
\end{eqnarray}
where $\epsilon_\pm$ are anticommuting spinor variables
with chirality $\gamma_5\epsilon_\pm=\pm\epsilon_\pm$.
   
The bosonic part of the action can be rewritten into more symmetric form
\begin{equation}
S_\phi\Rightarrow S_{\omega\varphi}
=\sum \Big[ \omega^\dagger \hat{D}\varphi+\varphi^\dagger
\hat{D}\omega\Big]+m\sum\varphi^\dagger\varphi+m\sum\omega^\dagger\omega,
\label{Somega}
\end{equation}
where $\omega$ and $\varphi$ are complex boson fields.
By integrating over $\omega(x)$ (or $\varphi(x)$), one can 
easily verify the equivalence of $S_\phi$ and $S_{\omega\varphi}$.

The lattice Dirac action (\ref{SD}) has exact chiral symmetry
for $M=0$.
This means that there appear multi-flavour Dirac fermions
with opposite chirality in the continuum limit, i.e., 
the species doubling.
$S_\chi$ with $M=0$ is therefore invariant under the following 
chiral $U(1)\times U(1)$ symmetry on the lattice,
\begin{eqnarray}
&& \chi_+(x) \rightarrow V_{\epsilon(x)}\chi_+(x),\;\;
\bar{\chi}_-(x)\rightarrow  \bar{\chi}_-(x)V^\ast_{-\epsilon(x)},
\nonumber  \\
&& \chi_-(x) \rightarrow W_{\epsilon(x)}\chi_-(x),\;\;
\bar{\chi}_+(x)\rightarrow  \bar{\chi}_+(x)W^\ast_{-\epsilon(x)},
\label{chiralS}
\end{eqnarray}
where $\epsilon(x)=(-)^{x_0+x_1}$ and $V_\pm, W_\pm \in U(1)$.
This symmetry plays a very important role in the discussion on the
phase structure as we shall see later on.

Expectation value of physical quantity $X$ is given by the following
functional integral,
\begin{equation}
\langle X \rangle= \int [DU D\bar{\chi}D\chi D\phi^\dagger D\phi]
\; P[U] \; e^{-S}\; X,
\label{physicalQ}
\end{equation}
where the probability distribution for the random vector potential is given by
\begin{equation}
P[U]=\exp \Big({N \over g} \sum_{x,\mu}
 Tr(U_\mu(x)+U^\dagger_\mu(x)) \Big),
\label{PU}
\end{equation}
$[DU]=\prod_{link}dU_\mu(x)$ is the Haar measure of $U(N)$
and $g$ is a parameter which controls the disorder strength.
As we explained in the introduction, the random vector potentials
are taken to be {\em compact} random variables.
This is in contrast with the previous studies where the vector
potential is taken to be {\em noncompact} and the probability
distribution is taken to be Gaussian.
Since in the weak-disorder case, i.e., the case of small $g$ , the
compact distribution (\ref{PU}) can be approximated as 
(here we show the Abelian case for notational simplicity)
\begin{eqnarray}
e^{{2N\over g}\cos A_\mu}, \;\; A_\mu\in [-\pi,\pi]
&\rightarrow& e^{-{N\over g}A^2_\mu+\cdots}, 
\; \; A_\mu \in (-\infty, +\infty)      \\
g&\rightarrow 0  \nonumber  
\end{eqnarray}
one may expect that the both compact and noncompact systems
give similar results at least qualitatively.\footnote{Actually 
this expectation is too naive. The compactness of the vector
potential plays a very important role for the correlation
functions in which singular configurations of $A_\mu$ give
dominant contribution. See discussion in Sect.5.} 
However in the strong-disorder case, substantial difference will appear.
Because of the regularization by the lattice,
topologically nontrivial configurations are all included
in the integral.
This is in contrast with the discussion in terms of the conformal 
field theory(CFT)\cite{mudry, caux}.

In the subsequent sections we shall perform $U(N)$-integral 
in (\ref{physicalQ}).
It is known that there are two ``phases" for this {\em one-link} integral, 
i.e.,
weak-coupling ``phase" for small $g$ and strong-coupling ``phase"
for large $g$.
Result of the integral exhibits a third-order phase transition
at a certain critical value of $g$\cite{BG}.
This is merely a matter of kinematics of the integral over the group 
$U(N)$ for large $N$.
However we think that there exists a genuine phase transition in the
system of the random Dirac fermions from weak to strong-disorder phases,
i.e., disorder-induced phase transition.


\setcounter{equation}{0}
\section{Weak-disorder regime}

Expectation values of physical quantities are given by 
(\ref{physicalQ}) and (\ref{PU}).
We shall first perform the functional integral of the vector
potential $U_\mu$.
Then let us consider the following one-link integral,
\begin{equation}
e^{W(\bar{D},D)}=\int dU_\mu \exp\Big[Tr(\bar{D}_\mu U_\mu
+U^\dagger_\mu D_\mu)\Big].
\label{onelink}
\end{equation}
In the present case, 
\begin{eqnarray}
D_\mu(x)^a_b&=& A_\mu(x)^a_b+{N\over g}\delta^a_b,  \nonumber  \\
\bar{D}_\mu(x)^a_b&=& \bar{A}_\mu(x)^a_b+{N\over g}\delta^a_b,  \nonumber  \\
A_\mu(x)^a_b&=&{\eta_\mu(x)\over 2}[\bar{\chi}_b(x+\mu)\chi^a(x)
+\omega^\dagger_b(x+\mu)\varphi^a(x)+
\varphi^\dagger_b(x+\mu)
\omega^a(x)],  \nonumber  \\
\bar{A}_\mu(x)^a_b&=&-{\eta_\mu(x)\over 2}[\bar{\chi}_b(x)\chi^a(x+\mu)
+\omega^\dagger_b(x)\varphi^a(x+\mu)+
\varphi^\dagger_b(x)
\omega^a(x+\mu)].
\label{DAA}
\end{eqnarray}
Let us introduce a parameter $s$ by
\begin{equation}
s={1 \over N} \sum^{N}_{a=1}x^{-1/2}_a=Tr(\bar{D}D)^{-1/2},
\label{sdef}
\end{equation}
where the $x_a$'s are eigenvalues of ${1\over N^2}\bar{D}D$.
In Ref.\cite{BG}, it is shown that there are two regimes for the
above integral (\ref{onelink}), i.e., weak-coupling regime 
for $s<2$ and strong-coupling regime for $s>2$.
For $D^a_b={N \over g}\delta^a_b$,
we can estimate the critical value of $g$ as 
$g_c=2$ from (\ref{sdef}).
In the present case there are extra factors $A$ and $\bar{A}$
in $D$ and $\bar{D}$ and then precise value of $g_c$ cannot
be determined.
However we can expect that for sufficiently small(large) $g$
the system is in the weak(strong)-coupling regime.

Let us consider the weak-disorder phase first, i.e., the phase of small $g$.
In this case the result of the $U(N)$-integral is given as\cite{BG} 
\begin{equation}
W(\bar{D},D)=N\Big\{2\sum_ax^{1/2}_a-{1\over 2N}\sum_{a,b}
\log (x^{1/2}_a+x^{1/2}_b)\Big\}.
\label{weakW}
\end{equation}
From (\ref{DAA})
\begin{equation}
\Big(\bar{D}_\mu D_\mu(x)\Big)^a_b=
    \Big( {N\over g}\Big)^2\delta^a_b+{N\over g}A^a_{\mu b}
+{N\over g}\bar{A}^a_{\mu b}+\bar{A}^a_{\mu c}A^c_{\mu b}.
\label{DD}
\end{equation}
It is straightforward to obtain $(\bar{D}D)^{1/2}$ in powers
of ${g \over N}$,
\begin{eqnarray}
(\bar{D}D)^{1/2,a}_b&=&{N\over g}\delta^a_b+{1\over 2}
(A^a_{\mu b}+\bar{A}^a_{\mu b})-{g\over 8N}(A^a_{\mu c}+\bar{A}^a_{\mu c})
(A^c_{\mu b}+\bar{A}^c_{\mu b})  \nonumber   \\
&& \; +{g \over 2N}\bar{A}^a_{\mu c}A^c_{\mu b}+O((g/N)^2).
\label{DD1/2}
\end{eqnarray}
Evaluation of the second term of the formula (\ref{weakW})
is also not so difficult.
As matrices $D$ and $\bar{D}$ have both large diagonal
matrix elements $N/g$, we can use the Taylor expansion for a regular
function $f(x,y)$,
\begin{eqnarray}
\sum_{a,b}f(x_a,x_b)&=&\sum_{a,b}f({1\over g^2}+w_a,
{1\over g^2}+w_b)   \nonumber   \\
&=&\sum_{n,m}{1\over n!m!}f^{(n,m)}({1\over g^2}, {1\over g^2})\;
Tr\Big({\bar{D}D \over N^2}
-{1 \over g^2}\Big)^n Tr\Big({\bar{D}D \over N^2}
-{1 \over g^2}\Big)^m,\nonumber
\end{eqnarray}
where $w_a$'s are eigenvalues of $({\bar{D}D \over N^2}-{1 \over g^2})$.
Leading-order terms are obtained as,
\begin{equation}
\sum_{a,b}\log (x^{1/2}_a+x^{1/2}_b)={g \over 2}(A^a_{\mu a}
+\bar{A}^a_{\mu a})+{g \over 2}{g\over N}\bar{A}^a_{\mu b}A^b_{\mu a}
+O((g/N)^2).
\label{sumlog}
\end{equation}

From (\ref{DD1/2}) and (\ref{sumlog}),
the effective theory which appears after the integration over the
$U(N)$ field for small $g$
is a SUSY extension of the Gross-Neveu
model.\footnote{More precisely, Dirac fermions in this theory has ``flavour"
degrees of freedom as a result of the species doubling.}
Detailed study of this model will be given elsewhere\cite{ichinose2}.
However here we mention that the four-Fermi coupling in the 
interaction term $\bar{A}^a_{\mu b}A^b_{\mu a}$ has the sign
which indicates instability of the Dirac fermion at the band center
or on criticality. 
By the usual $1/N$ expansion,
it is expected that the chiral condensation 
$\langle \bar{\psi}\psi \rangle \neq 0$
occurs in the present system. 
However the system under study is invariant under the chiral
$U(1)\times U(1)$ transformation (\ref{chiralS}), and therefore it is 
expected that only a quasi-long-range order exists as in the 
Kosterlitz-Thouless phase, though careful study is required
because of the existence of the bosons.
If this expectation is correct,
then excitations are massless ``pions", their SUSY 
partners and the Dirac fermions\cite{ichinose2}.

Closely related model appears in the case of 2D Dirac fermion in a random
{\em noncompact Abelian} vector potential, i.e., a SUSY Thirring model.
This is not surprising because in the weak-disorder limit
the compact measure of the vector potential is well approximated
by the noncompact Gaussian distribution, as we explained before.
In Ref.\cite{chamon}, it is shown that there exist an infinite
number of relevant operators with negative scaling dimensions.
Very recently Gurarie suggested a possible solution to this 
problem\cite{gurarie}.
In this argument, some ad hoc cutoff regularization is used
for the functional integral over the noncompact vector potential.
On the other hand in the present system, the compact Haar measure is
used for the integration over the vector potential and therefore
it is expected that a cutoff appears in a natural way.
We shall show that this is the case.
Relation between Gurarie's argument and the present
lattice model will be discussed later on.

Physical picture of the above phenomenon will be discussed rather in detail
in Sect.5 after investigation of the strong-coupling regime.
In the following section, we shall study the strong-disorder phase 
which is the main subject of the present paper.

\setcounter{equation}{0}
\section{Strong-disorder regime}

In the strong-coupling regime of the $U(N)$-integral,
$W(\bar{D},D)$ is given by the following formula\cite{BG},
\begin{eqnarray}
W(\bar{D},D)&=&N^2\Big\{ -{3\over 4} -c +{2 \over N}
\sum_a (c+x_a)^{1/2}  \nonumber  \\
&& \; \; -{1\over 2N^2}\sum_{a,b}\log (
(c+x_a)^{1/2}+(c+x_b)^{1/2})\Big\},
\label{strongW}
\end{eqnarray}
where $x_a$'s are again eigenvalues of ${1 \over N^2}\bar{D}D$
and a constant $c$ is given by 
\begin{equation}
1={1 \over 2N}\sum_a(c+x_a)^{-1/2}.
\label{c}
\end{equation}
As we explained above, the formula (\ref{strongW}) is suitable for 
large $g$.
The limit $g \rightarrow +\infty$ is nothing but the strong-coupling
limit of the SUSY lattice gauge theory which was 
studied in Ref.\cite{ichinose1}.
There we showed that the condensations like 
$\langle \bar{\chi}\chi\rangle$, $\langle {\varphi}^\dagger\varphi\rangle$,
etc. occur, whereas $\langle {\varphi}^\dagger\omega\rangle=
\langle {\omega}^\dagger\varphi\rangle=0$.
Here we assume a similar pattern of condensations for large $g$.
Then it is not so difficult to calculate the effective action
from (\ref{strongW}) by the $1/g$-expansion.
Here again the Taylor expansion is useful to convert the summation
over the eigenvalues $x_a$ into the trace of the matrix $\bar{D}D$.

After some calculation, we obtain\cite{ichinose3}
\begin{eqnarray}
{1\over N^2}W(\bar{D},D)&=&{1\over N}\Big[\sum_\pm F(\lambda_\pm)
-F(\xi)-F(\zeta) 
     \Big]+{\eta_\mu(x)\over gN^2}\Big[\chi^a(x)\bar{\chi}_a(x+\mu)
     G(\lambda)  \nonumber  \\
    && -\varphi^a(x){\omega}^\dagger_a(x+\mu)G(\xi)-\omega^a(x)
    {\varphi}^\dagger_a(x+\mu)G(\zeta)\Big]  \nonumber  \\
 && -{\eta_\mu(x)\over gN^2}\Big[\chi^a(x+\mu)\bar{\chi}_a(x)
 G(\lambda)  
    -\varphi^a(x+\mu){\omega}^\dagger_a(x)G(\xi) \nonumber  \\
   && - \omega^a(x+\mu){\varphi}^\dagger_a(x)
    G(\zeta)\Big] +O(1/(g^2N)),  
\label{Effstrong}
\end{eqnarray}       
where $\lambda_\pm$ etc are composite fields of $\chi_\pm$ etc,
and they are explicitly given by
\begin{eqnarray}
&& \lambda_\pm =\lambda_{\mu\pm}(x)=m_\pm(x)m_\pm(x+\mu), \;\; 
     m_\pm (x)={1\over N}\sum_a
     \chi^a_\pm(x)\bar{\chi}_{a\mp}(x),  \nonumber  \\
&& \xi= \xi_\mu(x)=\alpha(x)\beta(x+\mu), \;\;
\zeta=\zeta_\mu(x)=\beta(x)\alpha(x+\mu),   \nonumber \\
&&    \alpha(x)={1\over N}\sum_a\varphi^a(x){\varphi}^\dagger_a(x),  \;\;
   \beta(x)={1\over N}\sum_a\omega^a(x){\omega}^\dagger_a(x).
\label{composite}
\end{eqnarray}     
Functions $F(x)$ and $G(x)$ are given by    
\begin{eqnarray}
F(x)&=& 1-(1-x)^{1/2}+\log
[{1\over 2}(1+(1-x)^{1/2})],  \nonumber  \\
G(x) &=& (1+(1-x)^{1/2})^{-1}.  
\label{FG}
\end{eqnarray}
Actually there are additional terms of composites like 
$\chi^a(x){\varphi}^\dagger_a(x)$, but they do not have nonvanishing
expectation values and give only higher-order corrections in $1/N$
to the effective action of $m_\pm(x)$ etc. 

We expect that the $1/g$-expansion in (\ref{Effstrong})
has a finite convergence radius.
Then it is easily verified that the effective action can be 
written in terms of the composites $m(x), \alpha(x)$ and $\beta(x)$, 
or more precisely $\lambda_\mu(x), \xi_\mu(x)$ and 
$\zeta_\mu(x)$.\footnote{Strictly speaking, here we assume that 
the $U(N)$ symmetry is {\em not} spontaneously broken.}
Then we can introduce {\em elementary} fields
corresponding to the composite fields 
in the path-integral formalism.
First for the composite ``meson" field $m_\pm(x)$, we have identity like
(up to an irrelevant constant),
\begin{eqnarray}
Z^F_0(J)&\equiv& \int d\bar{\chi}d\chi e^{Jm}
=J^N=\int^{2\pi}_0 {d\theta \over 2\pi}
(\rho e^{i\theta})^{-N}\exp(J\rho e^{i\theta}) \nonumber  \\
&\equiv& \int d{\cal M}\; {\cal M}^{-N}\exp (J{\cal M}).
\label{mesonF}
\end{eqnarray}
Equation (\ref{mesonF}) means that the path integral of the
elementary meson fields ${\cal M}_\pm$ is defined 
by the above contour integral
and the radius $\rho$ should be taken for the angle integral 
to be well-defined, i.e., $\rho$ should be a {\em maximum} or {\em saddle
point} of the effective potential.

On the other hand for the boson-composite field $\alpha(x)$,
we can prove the following identity,
\begin{equation}
Z^B_0(J)\equiv\int d\bar{\varphi}d\varphi e^{-J\alpha}=J^{-N}
=\int^{+\infty}_{-\infty}d(\ln \Phi)\; \Phi^N  \exp(-J\Phi).
\label{Phi}
\end{equation}
In a similar way we introduce elementary field $\Psi(x)$ 
for the composite field $\beta(x)$.

From (\ref{strongW}), (\ref{Effstrong}) and (\ref{Phi}),
the effective action in the strong-disorder phase is obtained as
\begin{eqnarray}
{1\over N}S_{eff} &=& -\sum_{x,\mu,\pm}\Big[F(\lambda_{\mu\pm}(x))
-{1\over 4}\log \lambda_{\mu\pm}(x)\Big] 
-M\sum_x({\cal M}_+-{\cal M}_-) \nonumber  \\
&&\;\; +\sum_{x,\mu}\Big[F(\xi_{\mu}(x))
-{1\over 4}\log \xi_{\mu}(x)\Big] +m\sum_x\alpha  \nonumber  \\
&& \;\; +\sum_{x,\mu}\Big[F(\zeta_{\mu}(x))
-{1\over 4}\log \zeta_{\mu}(x)\Big] +m\sum_x\beta +O(1/g^2),
\label{Seff2}
\end{eqnarray}
where $\lambda_{\mu\pm}(x)={\cal M}_\pm(x){\cal M}_\pm(x+\mu)$,
$\xi_\mu(x)=\Phi(x)\Psi(x+\mu)$ and $\zeta_\mu(x)=\Psi(x)\Phi(x+\mu)$.
Terms of $O(1/g^2)$ have a similar structure to the leading-order terms.

Then it is  straightforward to study the structure of the
ground state and low-energy excitations.
For vanishing masses $M=m=0$, the ground state is parameterized as
follows,
\begin{equation}
\langle {\cal M}_\pm (x)\rangle =\left\{
  \begin{array}{ll}
  vU_{0\pm}, & \;\; \mbox{at even sites}  \\
  vU^\ast_{0\pm}, & \;\; \mbox{at odd sites}
  \end{array}\right.
\label{ground1}
\end{equation}
 \begin{equation}
\langle \Phi (x)\rangle =\left\{
  \begin{array}{ll}
  ve^{\sigma_1}, & \;\; \mbox{at even sites}  \\
  ve^{-\sigma_2}, & \;\; \mbox{at odd sites}
  \end{array}\right.
\label{ground2}
\end{equation}   
 \begin{equation}
\langle \Psi (x)\rangle =\left\{
  \begin{array}{ll}
  ve^{\sigma_2}, & \;\; \mbox{at even sites}  \\
  ve^{-\sigma_1}, & \;\; \mbox{at odd sites}
  \end{array}\right.
\label{ground3}
\end{equation}     
where $U_{0\pm} \in U(1)$, $\sigma_i$'s $(i=1,2)$ are real numbers and 
$v$ is obtained from the stationary
condition of the effective potential,
\begin{equation}
{dV(v^2)\over dv^2}={dF(v^2)\over dv^2}-{1\over 4v^2}=0,
\label{stationary}
\end{equation}
with the following solution
\begin{equation}
v^2=3/4.
\label{v}
\end{equation}
The mass terms lift the above degeneracy and 
determine the expectation values as $\langle {\cal M}_+\rangle
=-\langle {\cal M}_-\rangle$ and $\sigma_1=\sigma_2=0$.
Obviously the ground state preserves the SUSY\cite{ichinose1}.

For vanishing masses, there exist the degeneracies of the ground state
parameterized by $U_{0\pm}$ and $\sigma_i$ which originate from the chiral
symmetry (\ref{chiralS}) and its SUSY counterpart for $\varphi$ and 
$\omega$.
By the Coleman-Mermin-Wagner theorem, in two dimensions
continous symmetry is not spontaneously
broken and there exists no long-range order.
Therefore we cannot expect the condensations $\langle{\cal M}_\pm
\rangle \neq 0$, etc.
Instead the ground state exists in the Kosterlitz-Thouless phase
with gapless excitations.
In fact we can explicitly show the existence of massless modes which
destroy the off-diagonal long-range order.
These excitations are described by the ``pion" fields,
\begin{equation}
{\cal M}_\pm(x) = 
       \left\{
  \begin{array}{ll}
  vU_{\pm}(x)=ve^{i\pi_\pm(x)}, & \;\; \mbox{at even sites}  \\
  vU^\ast_{\pm}(x)=ve^{-i\pi_\pm(x)}, & \;\; \mbox{at odd sites}
  \end{array}\right.
\label{pionF}
\end{equation}  
and similar SUSY excitations for $\Phi(x)$ and $\Psi(x)$.
From (\ref{ground2}) and (\ref{ground3}), it is obvious that 
these low-energy
excitations are nothing but ``density wave" of the SUSY bosons
which is commensurate with the lattice structure.

Effective action of $\pi_\pm(x)$ is obtained as follows
from (\ref{Seff2}) and (\ref{pionF}),
\begin{eqnarray}
S_\pi&=&{N \over 2}C\sum [\pi_\pm(x+\mu)-\pi_\pm(x)]^2, \nonumber  \\
C&=& F''(v)v^4+{1\over 4}, \;\; v^2=3/4.
\label{Spi}
\end{eqnarray}
Therefore the correlator of ${\cal M}_\pm(x)$ exhibits a power-law
decay
\begin{equation}
\langle \bar{\psi}\psi(x)\bar{\psi}\psi(0)\rangle
=\langle {\cal M}_\pm(x){\cal M}_\pm(0)\rangle 
\sim  |x|^{-1/(2\pi NC)}.
\label{correlation}
\end{equation}
Corrections of $O(1/g^2)$ can be calculated systematically and the 
scaling dimension of ${\cal M_\pm}$ acquires correction of $O(1/g^2)$. 

There appear no signs of instability of the ground state.
Low-energy excitations in the boson sector are given by
local fluctuations of $\sigma_i$ $(i=1,2)$ in Eqs.(\ref{ground2}) 
and (\ref{ground3}).
They are SUSY counterparts of $\pi_\pm(x)$ and stable.
All correlation functions in the fermion sector have 
nonsingular behaviour like Eq.(\ref{correlation}).
This is in sharp contrast with the previous 
results which show that fermion composite operators
(as well as boson composite operators)
have negative scaling dimensions\cite{ludwig,chamon,mudry}.
We think that the compactness of the functional-integral measure
of the vector potential plays a very important role for the stability.
This important point will be discussed in the following section.

From the discussion given so far, it is obvious that 
the algebraic decay of the correlation functions in the 
strong-disorder phase comes from
the exact chiral symmetry on the lattice which is a result 
of the species doubling.
In the single-flavour case in the continuum, there exists anomaly
in the chiral symmetry because of the coupling with the vector potential,
and therefore the genuine condensation $\langle\bar{\psi}\psi\rangle
\neq 0$ is possible even in two dimensions just as
in the Schwinger model.\footnote{In other words, the would-be 
Nambu-Goldstone boson acquires a mass by the chiral anomaly,
and it generates no severe infrared singularities.}

\setcounter{equation}{0}

\section{Discussion}

In this paper we studied Dirac fermions in a random $U(N)$
vector potential.
We employed the lattice regularization and the compact Haar measure
in order to make the
functional integral over $U(N)$ vector potential well-defined.
In this formalism topologically nontrivial configurations are
all integrated over.
This is in sharp contrast with the approaches given so far.
The ensemble average over the random vector potential was taken
by introducing bosons in a SUSY way.
We think that this approach is important because there appeared
some evidences that there exists a disorder-induced phase transition
in the present system.
In order to investigate this problem, a well-defined formalism
is indispensable.

For the one-link $U(N)$ integral, it is known that there are
two regimes, i.e., the weak and strong-coupling regimes,
which correspond to the weak and strong-disorder cases, respectively.
We obtained effective theory by integrating over the vector potential
in both regimes.
In the weak-disorder phase, the effective theory is a SUSY 
extension of the Gross-Neveu model.
We call this regime phase A.
Detailed studies on the effective field theory of 
the phase A will be reported elsewhere,
but sign of the effective coupling constant of the four-Fermi 
interaction indicates
instability of the ground state to the state with the chiral condensation.
From the investigation of the Gross-Neveu model by the $1/N$ expansion,
we expect that there appears the chiral condensation 
with a quasi-long-range order.

Then we studied the strong-disorder phase rather in detail.
We call this regime phase B.
We showed that in the phase B the density operator of the fermion 
has the quasi-long-range order and low-energy excitations are the
``pions" and density wave of the bosons whereas no Dirac fermions
with the original $U(N)$ quantum number appear.
This result stems from not only the strong-disorder properties
of the vector potential but also
the exact chiral symmetry and its SUSY
counterpart on the lattice, i.e., anomaly cancellation by the species
doubling.
Therefore, the result indicates that genuine condensation
of the fermion density operator occurs in the single-flavour case
with chiral anomaly just as in the Schwinger model.
Anyway, we expect the existence of the phase transition
at a certain critical value of the disorder strength $g_c$
from the phase A to B.\footnote{Order of this phase transition
can be of third-order as the one-link integral indicates.}

Let us discuss physical picture of the above phenomena.
Coupling with the vector potential reduces
the effective hopping of fermions which is simply given by
$t_{\mbox{\small eff}}=t\cdot\langle U_\mu (x)\rangle$ 
where $t$ is the original
hopping parameter.
Obviously as increasing the disorder strength $g$, $t_{\mbox{\small eff}}$
decreases and fermionic states tend to localize.
Study in this paper shows that in the present system that localization
phenomenon occurs with the chiral
condensation of fermions which makes the fermions massive.
As $g$ is increased further more, even a local hopping of a single fermion
cannot occur anymore and movement of a fermion always accompanies the same
movement of an anti-fermion for fluctuations of the vector
potential in the hopping
cancels out with each other, i.e., $\langle U_\mu(x)\rangle=0$ but
$\langle U_\mu(x) U^\dagger_\mu(x)
\rangle =1$.
That is, only a ``bound state" of fermion and anti-fermion pair can 
move in this phase.
Then condensation of fermion density operator is generated.
This phase (the phase B) is more or less similar to the conventional 
confinement
phase of the strong-coupling gauge theory.
From this picture, disorder-induced phase transition is naturally
understood.

Finally let us discuss relation between the results in this paper
and previous studies.
For both the Abelian and non-Abelian cases, it is known that there 
exist an infinite number of relevant operators with negative dimensions
if the noncompact Gaussian distribution is used for
the vector potential.
It indicates some instability of the critical line.
On the other hand in the present study, no signs of instability appear
at least in the effective action obtained by integrating over the 
vector potential.\footnote{Strictly speaking, the correlation function
in the boson sector $\langle\Phi(x)\Psi(0)\rangle$ tends to
diverge for $|x| \rightarrow \infty$.
However correlators in the fermion sector, which are physical quantities
in the present system, exhibit no singular behaviour.}
We think that this stability stems from the compactness of the
integral measure of the vector potential.
Actually as recently Gurarie discussed\cite{gurarie} and the 
study on the random XY model shows\cite{XY2}, the instability
comes from the noncompactness of the vector potential.
More explicitly in the discussion in the continuum, the vector
potential $A_\mu$ is parameterized as follows 
(we here consider the Abelian case for simplicity\footnote{For
the Abelian case, similar discussion of the strong-disorder case is possible.
See Ref.\cite{ichinose4}.}),
\begin{equation}
A_\mu=\epsilon_{\mu\nu}\partial_\nu \theta+\partial_\mu \eta,
\;\; \epsilon_{01}=-\epsilon_{10},
\label{vecp}
\end{equation}
where $\theta(x)$ and $\eta(x)$ are scalar fields and $\theta(x), 
\eta(x)\in 
(-\infty,+\infty)$.
The instability and the negative dimensions of the relevant operators
essentially come from the following correlation function of 
$\theta(x)$\cite{XY2},
\begin{equation}
\langle e^{-c\theta(x_1)}e^{c\theta(x_2)}\rangle,
\label{instability}
\end{equation}
where $c$ is a real number and the above expectation value is
evaluated with the following probability distribution,
\begin{equation}
P[\theta]\propto \exp\Big\{-{1\over g}\int d^2x \; (\partial_\mu
\theta)^2\Big\}.
\end{equation}
Then it is not difficult to show that the operator $e^{c\theta(x)}$
has a negative dimension and the correlator in Eq.(\ref{instability})
tends to diverge for large $|x_1-x_2|$.
However it is obvious that if the integral region of $\theta(x)$ is 
compact, this divergence does not occur.
This is an essential point of Gurarie's argument\cite{gurarie}.

Gurarie used some ad hoc cutoff regularization for 
the functional space of $\theta(x)$.
On the other hand in this paper, we use the Haar measure which is compact,
and therefore we expect that a similar cut off appears naturally
in the present formalism.
Actually we can parameterize the vector potentials $U_\mu(x)$ 
in terms of two $U(1)$ fields $u(x)$ and $v(x+{\hat{0}+\hat{1} \over 2})$
($\hat{0}(\hat{1})$ is the unit vector of the $0(1)$ direction),
where $u(x)$ is defined on the sites of the original lattice and 
$v(x+{\hat{0}+\hat{1} \over 2})$ is on the sites of the {\em dual} lattice,
\begin{equation}
U_0(x)=u(x+\hat{0})u^\ast (x)v(x+{\hat{0}+\hat{1}\over 2})
v^\ast(x+{\hat{0}-\hat{1}\over 2}),
\label{vecp2}
\end{equation}
and similarly for $U_1(x)$ where $u^\ast(x)$ is the complex
conjugate to $u(x)$ etc.
It is not difficult to show that if we impose the conditions like
$\prod_xu=\prod_xv=1$, then there is no ambiguity in 
this parameterization.
It is obvious that $\theta(x)(\eta(x))$ in the continuum 
expression (\ref{vecp})
is related to $v(x+{\hat{0}+\hat{1} \over 2})(u(x))$ in (\ref{vecp2}) 
as follows,
\begin{equation}
\theta\sim\ln v, \;\; \eta\sim\ln u.
\end{equation}
Therefore the integral region of $\theta(x)$ is compact in the 
present formalism, $\theta(x) \in [-\pi, +\pi]$, i.e., there
exists the natural cutoff.
Please notice that this compact region of $\theta(x)$ remains
the same even if we recover the lattice spacing $a_L$.

In the sense explained above, the model in the present paper
is {\em different} from those with noncompact random vector potential
which were studied in the previous papers.
Our model is close to the network models by Chalker and 
Coddington\cite{network} and their field-theory models\cite{lee}.
Then it is not so surprising that the stable ground state appears
and the low-energy excitations are the massless ``pions" etc
as in the Kosterlitz-Thouless phase even at the strong-disorder limit.
Also we can conclude that
the disorder-induced phase transition which we found in this paper
is a new one.
This phase transition is expected to be of a topological nature
of the vector potential but it is not definitive at this stage.

\vskip 2cm
{\Large{\bf Acknowledgements}}

The author is grateful to H.Mukaida for useful discussion.\\  
\vskip 2cm
{\bf Note added}

After submitting this paper, we got acquainted with the paper
by Altland and Simons\cite{altland} which also studies the random
flux model on a lattice.


\end{document}